\documentclass[prd,preprintnumbers,nofootinbib]{revtex4} 
\usepackage{graphicx} 
\usepackage{amsmath}
\usepackage{amsfonts,amsbsy}
\usepackage{amssymb}
\usepackage{appendix}
\usepackage{slashed}

\def\be{\begin{equation}}
\def\ee{\end{equation}}
\def\bea{\begin{eqnarray}}
\def\eea{\end{eqnarray}}

\def\Tr{{\rm Tr}}
\def\eq#1{{Eq.~(\ref{#1})}}
\def\fig#1{{Fig.~\ref{#1}}}
\def\fig#1{{Fig.~\ref{#1}}}
\def\s#1{{\slashed{#1}}}

\def\k{{\mathbf k}}

\def\p{{\mathbf p}}
\def\r{{\mathbf r}}
\def\b{{\mathbf b}}
\def\q{{\mathbf q}}
\def\k{{\mathbf k}}

\def\x{{\mathbf x}}
\def\y{{\mathbf y}}

\begin{document}

\title{\bf DPS in CGC: HBT correlations in double inclusive photon production }


\author{Alex Kovner$^{1}$ and Amir H. Rezaeian$^{2,3}$}

\affiliation{
$^1$ Dept. of Physics, University of Connecticut, High, Storrs, CT 06269, USA\\
$^2$ Departamento de F\'\i sica, Universidad T\'ecnica
Federico Santa Mar\'\i a, Avda. Espa\~na 1680,
Casilla 110-V, Valparaiso, Chile\\
$^3$  Centro Cient\'\i fico Tecnol\'ogico de Valpara\'\i so (CCTVal), Universidad T\'ecnica
Federico Santa Mar\'\i a, Casilla 110-V, Valpara\'\i so, Chile
}

\begin{abstract}
We introduce a technique to study double parton scattering (DPS) in the Color-Glass-Condensate (CGC) approach. We show that the cross-section of the DPS in the CGC approach is calculable in terms of new nonperturbative objects,  generalized double transverse momentum-dependent
parton distribution (2GTMD) functions. We investigate the production of pairs of prompt photons from two partons in the projectile hadron in high-energy proton-nucleus collisions.  We show that even for independent partons in the projectile,  the prompt photon correlation function exhibits  Hanbury Brown and Twiss (HBT) correlations.  The width of the HBT peak is controlled by the  transverse distance between the parton of the pair, which is of the order of the proton size. Thus, the HBT measurements in two-particle production such as prompt photon pairs provide useful information about the nonperturbative 2GTMDs.  
 
\end{abstract}

\maketitle

\section{Introduction}
The nature of long range rapidity correlations in hadron production observed in proton-proton (p-p) and proton(deuteron)-nucleus (p-A) collisions at the LHC and RHIC has been a subject of intense investigation during the last several years \cite{exp-r1,exp-r2,exp-r3,exp-r4,exp-r5}.  The big question to be answered is whether these correlations arise due to strong collective effects in final state interactions, or due to quasi collectivity present in the initial state wave function which is imprinted on the spectrum of produced particles.

\begin{figure}[t!]                                       
                                  \includegraphics[width=9 cm] {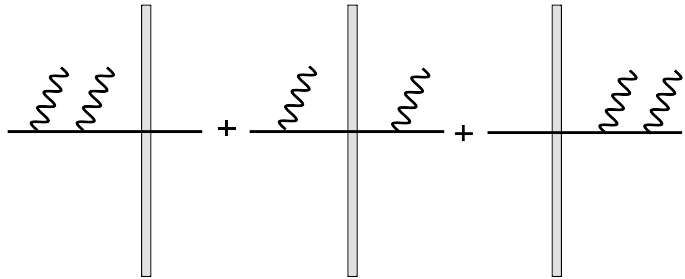}             
\caption{The diagrams contributing to two prompt photons production from one quark in the background of the CGC field. The shaded box (the CGC shock waive) denotes the interaction of a quark to all orders with the background field via multiple gluon exchanges.   }
\label{f0}
\end{figure}

\begin{figure}[t!]                                       
                                  \includegraphics[width=12 cm] {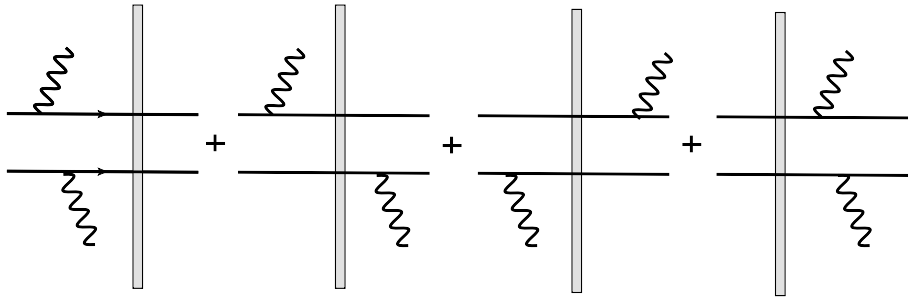}             
\caption{The diagrams contributing to production of two prompt photons from two quarks in the background of the CGC field. The notation is the same as in \fig{f0}.}
\label{f1}
\end{figure}

Since first principle analysis of hadron production in a dense environment is very hard, it makes sense to look at simpler probes of this system. Prompt photons have been one such probe that has been used to probe the putative quark-gluon-plasma state created at the early stages of heavy ion collisions \cite{boris-photon,pho-cgc1,pho-cgc2,amir-photon1,amir-photon2,dy-ana,amir-photon3,pho-all,ncgc1,ncgc2,ncgc3}. Since photon interactions are weak, the correlations between emitted photons, if such exist would most certainly probe the structure of initial state alone, and it is interesting to see what can be learned from it. This was the motivation of our previous papers on the subject \cite{kr}. In Ref.\,\cite{kr} we considered production of two photons from the same quark in reaction of the type (shown in \fig{f0}), 
\be
q+A\to \gamma(k_1) +\gamma(k_2) + \text{jet}(q) + X,  \label{2pg-old}
\ee
and have found an interesting correlated structure albeit short range in rapidity compared to the di-hadron correlations.  Here, we consider production of two photons from two valence quarks, i.e. process of the type (shown in \fig{f1}),
\be
q+q+A\to \gamma(k_1) +\gamma(k_2) + \text{jet}(q) + \text{jet}(q^\prime)+ X. \label{2pg-new}
\ee
Naively one might think that such independent emission process does not lead to correlations in double photon production. However this is not necessarily the case. Since photons are bosons, upon further reflection one expects to see the Hanbury Brown and Twiss (HBT) correlations between photons emitted from two independent sources. Such correlations involving gluons were discussed in the context of hadron production in p-A scattering in the CGC approach in \cite{hbthadrons}. In principle they result in a peak for production of same sign transverse momentum pairs, with the radius of correlation in momentum space given by the inverse gluonic radius of the proton. The hadron HBT signal is however rather fragile and is easily masked by final state effects.  One expects the photon HBT to be much more resilient. One of the main purpose in this paper is to qualitatively study this effect.

Although our main interest  and explicit calculations are geared towards di-photons production, the approach itself is more general and can be applied to any process of the Double Parton Scattering (DPS) type (given in \eq{2p-new} below). 

The topic of multiparton interactions is one of the most important focal points of studying the multi-particle correlations in pQCD. In the CGC approach, at leading order, the so-called single parton scattering (SPS) processes in the pQCD framework, correspond to a single parton scattering to the CGC shock wave. The CGC shock wave includes the interaction of a parton to all orders with the background color field of the target. Therefore, in the CGC approach, the corresponding SPS contribution for two-particle production is obtained by considering the following process  
\be
\text{parton}+A\to \text{particle}(k_1) +\text{particle}(k_2)  + X. \label{sps}
\ee
Effects of saturation on di-hadron correlations originating from SPS  have been studied  in \cite{dihadron}.
A potentially ``richer'' source of correlations are processes where two partons in one projectile hadron collide with the CGC shock wave. Such a process in the language of pQCD, is the so-called double parton scattering (DPS), 
\be
\text{parton}+\text{parton}+A\to \text{particle}(k_1) +\text{particle}(k_2) + X. \label{2p-new}
\ee
In the CGC framework  processes where two observed particles originate from different sources of the color field were studied in the soft limit in 
Refs.\,\cite{ridge0,ridge1,ridge2,ridge3,ridge1-raju,ridge33,ridge4,ridge5,hbthadrons}. While for SPS, the CGC and pQCD approaches are conveniently bridged with the help of the hybrid formalism \cite{dj-rhic}, such connection has not been made for DPS so far.

In the present paper we extend the hybrid formalism to include the DPS in the CGC approach. We show that the cross-section of the DPS in the CGC is calculable in terms of new nonperturbative objects, the generalized double transverse momentum-dependent parton distribution (2GTMD) functions. In the context of the collinear factorization, a similar object, the so-called generalized double parton distribution (2GPD) appears in studies of the DPS \cite{marke}.  We propose that the properties of the 2GTMDs can be studied in the small-x kinematics within this hybrid CGC approach. In particular we show that the di-photon HBT correlations are naturally express in terms of the di-quark 2GTMDs. We also point out that di-hadron correlations at high energy (in forward direction) should be sensitive to di-gluon 2GTMD, and such processes should be included as corrections to the calculations of Ref.\,\cite{dihadron}.

We will be working within a variant of the "hybrid" approximation \cite{dj-rhic} which is appropriate for forward photon production. In the hybrid  CGC approach, we assume that the small-x gluon modes of the nucleus have a large occupation number so that the target nucleus can be described in terms of a classical color field.  This should be a good approximation for large enough nucleus at high-energy\footnote{Note that there is growing evidence that supports the idea that a proton at very high energy and especially at very forward rapidity can be considered as a dense system as well and therefore in principle the same approximation also applies to high energy p-p scatterings, see for example Refs.~\cite{jav1,amir-hera1,amir-hera2,na-sat,kn-rhic,rhic-cgc,pp-LR,jav-d,pa-raju,jav-pa,pa-R,aa-LR,raju-glasma,hic-ap,pa-jam}.}. This color field emerges from the classical Yang-Mills equation with a source term provided by faster partons. The renormalization group equations which govern the separation between the soft and hard models are then given by the non-linear Jalilian-Marian, Iancu, McLerran, Weigert, Leonidov, Kovner (JIMWLK) evolution equations \cite{jimwlk} (see below).  We further assume that the projectile proton is in the dilute regime and can be described in ordinary perturbative approach. The process \eq{2p-new} involves double parton scattering, and therefore standard collinear  parton distribution functions are not sufficient to characterize the incoming proton state. We will therefore need to model the proton structure in a slightly more refined way.

In the following we will derive the di-photon cross-section starting from some simple and intuitive assumptions about the wave function of the two incoming quarks inside the projectile hadron.
We show that at large $N_c$ within the CGC approach, the cross-section is determined by the dipole scattering amplitude.  We also show that within the standard collinear factorization approach the HBT peak has zero width. This is not at all surprising, since the width is expected to be of the order of inverse proton size, while in collinear factorization this size is effectively infinite. Thus any realistic study requires us to go beyond the standard collinear factorization.  Since the basic process we consider involves double parton scattering, our final expressions requires double parton distributions.  These objects are not well determined experimentally, and therefore we do not attempt detailed quantitative predictions. Instead we limit ourselves to qualitative analysis based on a simple model of the initial wave function. We show that the width of the HBT peak is indeed given by the inverse size of this distribution in coordinate space. Therefore, the HBT measurements for two-particle production such as prompt photon pairs provide useful information about the nonperturbative 2GTMDs.

This paper is organized as follows: In Sec. II, we first provide a concise description of theoretical framework for calculating the DPS contribution in the CGC approach. As an example, we focus on calculating the cross-section of a pair of prompt photon and a pair of jet in high-energy p-A collisions.  In Sec. III, we present our results for inclusive prompt di-photon production obtained from the DPS contribution in the CGC approach. We will also discuss Hanbury Brown and Twiss correlations for di-photon production in high-energy p-p and p-A collisions. We summarize our main results in Sec. IV.

\section{Semi-inclusive diphoton+dijet production in proton-nucleus collisions}
In this section, we present the basics of computation of the cross-section for the process given in \eq{2p-new}. 
Although, our formulation here is valid for the general production given in \eq{2p-new}, in the following we focus on a case where two  produced  particles are prompt photons,
\be
q(p_1)+q(p_2)+A\to \gamma(k_1) +\gamma(k_2) + \text{jet}(q) + \text{jet}(q^\prime)+ X.  \label{2-ph}
\ee
We consider the leading order approximation in a dilute-dense collisions at forward rapidity, for example in proton-proton  or/and  in proton-nucleus collisions. In this setup, the  two valence quarks from 
the projectile wave function  emit two photons via Bremsstrahlung. The {\it two}  photon+jet systems are put on shell by interacting coherently over the whole longitudinal extent of the target, see \fig{f1}. Although the scattering of the two quarks is independent, the production of two  photon+jet systems  is not independent due to the interference diagrams (shown in \fig{f1}). 

In the following, two-dimensional vectors in transverse space are written in  boldface. 

The cross-section for production of two quarks with momentum $q$ and $q^\prime$ and two
prompt photons with momenta $k_1$ and $k_2$ in the scattering of two on-shell quarks with momentum $p_1$ and $p_2$ off a hadronic
target (either a proton or a nucleus), given  in \eq{2-ph} can be written in the following general form, 
\be d\, \sigma^{qq\to \gamma \gamma qq} = \frac{d^3 k_1}{(2\pi)^3\,2 k_1^-}  \frac{d^3 k_2}{(2\pi)^3\,2 k_2^-} \frac{d^3 q}{(2\pi)^3\, 2
  q^-} \frac{d^3 q^\prime}{(2\pi)^3\, 2
  q^{\prime -}} \,\frac{1}{4p^{-}_1p^{-}_2}\,\ \langle| \langle \text{jet}(q), \text{jet}(q^\prime), \gamma(k_1),\gamma(k_2)| \text{Proton} \rangle|^2 \rangle_{\text{color sources}}. 
\label{m1}
\ee
For explicit calculations one needs the two quarks distribution in the proton wave function. In full generality we can write
\be
|\text{Proton} \rangle=\frac{1}{2N_c}\sum_{s_1,s_2, c_1,c_2}\int \int  \frac{d^3p_1}{(2\pi)^3}\frac{d^3p_2}{(2\pi)^3} \sum_X\tilde{\mathcal{A}}(p_1,p_2,s_1,s_2,c_1,c_2;X)|p_1,s_1,c_1; p_2,s_2,c_2, X\rangle,
\ee
where $(s_1,s_2)$ and $(c_1,c_2)$ are the spin and the color indices of two quarks (in the projectile proton) respectively, and for simplicity we have assumed that the two quarks have the same flavor.   The generalization to include $u$ and $d$ quarks is straightforward, and would result in promoting $\tilde{\mathcal{A}}$ to a matrix in flavor space. In this expression  $X$ stands for all the "spectator" degrees of freedom in the proton wave function which are integrated over inclusively in the process Eq. (\ref{2-ph}). These include the occupation numbers of the spectator quarks and gluons as well as the momentum, spin and color index of these spectators.
The factor $\frac{1}{2N_c}$ was introduced for future convenience.

To calculate the cross section we require the reduced two quark density matrix
\be\label{density}
 \sum_X |\text{Proton} \rangle\langle \text{Proton}|=\int_{p_1,p_2,p'_1,p'_2}\tilde{\mathcal{R}}(p_1,p_2,p'_1,p'_2,s_1,s_2,s'_1,s'_2,c_1,c_2,c'_1,c'_2)|p_1,s_1,c_1; p_2,s_2,c_2\rangle\langle p'_1,s'_1,c'_1; p'_2,s'_2,c'_2|, 
 \ee
 where
 \be
 \tilde{\mathcal{R}}(p_1,p_2,p'_1,p'_2,s_1,s_2,s'_1,s'_2,c_1,c_2,c'_1,c'_2)=\sum_X \tilde{\mathcal{A}}(p_1,p_2,s_1,s_2,c_1,c_2;X)\tilde{\mathcal{A^*}}(p'_1,p'_2,s'_1,s'_2,c'_1,c'_2;X). 
 \ee
While it is possible to perform the calculations with the general density matrix Eq. (\ref{density}), for simplicity we will assume that the integration over the spectator partons leads to  decorrelation  of spin and color of the two active quarks in the density matrix. We will also take a simple product ansatz for the density matrix in momentum space. In other words we take the following simple form 
\be \label{product}
\tilde{\mathcal{R}}(p_1,p_2,p'_1,p'_2,s_1,s_2,s'_1,s'_2,c_1,c_2,c'_1,c'_2)=\left(\frac{1}{2N_c}\right)^2\tilde{\mathcal{P}}(p_1,p_2)\tilde{\mathcal{P}}^*(p'_1,p'_2). 
\ee
 The function $ \tilde{\mathcal{P}}(p_1,p_2)$ now  determines the distribution of the two quarks in the proton on the amplitude level. 
 
We  stress that in practical terms the product ansatz makes very little difference  since we are not going to assume that the  momentum of the pair $p_1+p_2$ is equal to the total momentum of the proton. The most important feature of Eq. (\ref{product}) is that the two quarks in \eq{product} are taken to be totally uncorrelated in spin and color. One can check explicitly, that taking an analogous factorized form  for a single quark density matrix reproduces exactly the standard expressions for cross-section from a single quark where one averages over spin and color on the cross-section level, and the parton distribution function (pdf) given by $\int_{p_T}|\tilde{\mathcal{P}}(p_T,x)|^2$.  We will quote the result obtained for a general density matrix later, see Eq.(\ref{dens}).

With this simple form of the reduced density matrix our calculation amounts to replacing the proton wave function in Eq. (\ref{m1}) by
\be\label{proton}
 |\text{Proton} \rangle\rightarrow |\text{Two quarks}\rangle=\frac{1}{2N_c}\sum_{s_1,s_2, c_1,c_2}\int \int  \frac{d^3p_1}{(2\pi)^3}\frac{d^3p_2}{(2\pi)^3} \tilde{\mathcal{P}}(p_1,p_2)|p_1,s_1,c_1; p_2,s_2,c_2\rangle. 
\ee

\begin{figure}[t]                                       
                                  \includegraphics[width=9 cm] {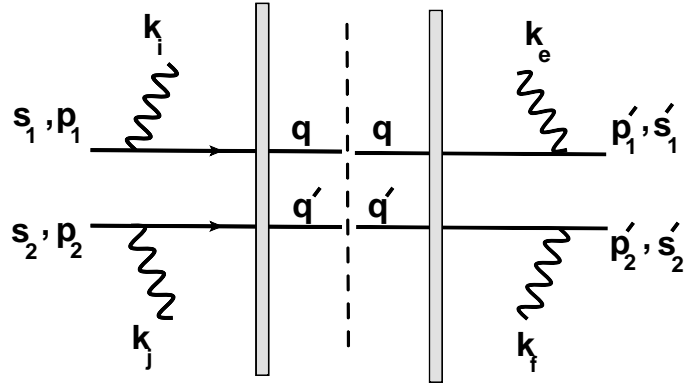}             
\caption{A typical diagram contributing to two prompt photons production from two quarks in the background of the CGC field. The diagrams on the left and right side of the dashed line correspond to the amplitude and the complex conjugate amplitude.  The cross-section at LO is given by the sum of four diagrams of this type shown in \fig{f1}.   }
\label{f2}
\end{figure}

In the following, the spin and the color indices of two quarks in the conjugate amplitude  are denoted by $(s_1^\prime,s_2^\prime)$ and $(c_1^\prime,c_2^\prime)$, respectively, see \fig{f2}. 
The indices $(s, s^\prime)$ and $(c, c^\prime)$ denote spin and color indices of the produced two quarks in the final state. The matrix element of the scattering amplitude in \eq{m1} is given by, 
\bea 
&&\langle \text{jet}(q), \text{jet}(q^\prime), \gamma(k_1),\gamma(k_2)| \text{two quarks} \rangle =\frac{1}{2N_c} \sum_{s,s^\prime,s_1,s_2,c,c^\prime,c_1,c_2} \int \int \frac{d^3p_1}{(2\pi)^3}\frac{d^3p_2}{(2\pi)^3}\tilde{\mathcal{P}}(p_1,p_2)  \nonumber\\
&&\Big[ \langle q, s, c; k_1|p_1,s_1, c_1\rangle \langle q^\prime, s^\prime, c^\prime, k_2|p_2, s_2, c_2\rangle 
+  \langle q, s, c; k_2|p_1,s_1, c_1\rangle \langle q^\prime,s^\prime, c^\prime; k_1|p_2, s_2, c_2\rangle \nonumber\\
&&+
\langle q^\prime, s^\prime, c^\prime; k_1|p_1,s_1, c_1\rangle \langle q,s, c; k_2|p_2, s_2, c_2\rangle 
+
\langle q^\prime, s^\prime, c^\prime; k_2|p_1,s_1, c_1\rangle \langle q,s, c; k_1|p_2, s_2, c_2\rangle
\Big]. \
\label{m2}
\eea
In \eq{m2}, we perform the sum over the spin and the color of produced quarks.   
Here  $ \langle q, s, c; k_1|p_1,s_1, c_1\rangle $ is the perturbative  production amplitude for the process,
\be
q(p_1,s_1,c_1)+A\to \gamma(k_1) + q(q,s,c) + \text{jet}(q^\prime)+ X. 
\ee
 For brevity of notation, the photon polarization indices, and summation over photon polarization are implicit in \eq{m2} and throughout this paper. The expression in \eq{m2} can be simplified by rearranging terms, 
\bea 
&&\langle \text{jet}(q), \text{jet}(q^\prime), \gamma(k_1),\gamma(k_2)| \text{two quarks} \rangle =\frac{1}{2N_c} \sum_{s,s^\prime,s_1,s_2,c,c^\prime,c_1,c_2} \int \int \frac{d^3p_1}{(2\pi)^3}\frac{d^3p_2}{(2\pi)^3} \mathcal{P}(p_1,p_2)  \nonumber\\
&\times& \Big[\langle q, s, c; k_1|p_1,s_1, c_1\rangle \langle q^\prime, s^\prime, c^\prime, k_2|p_2, s_2, c_2\rangle  + \langle q, s, c; k_2|p_1,s_1, c_1\rangle \langle q^\prime, s^\prime, c^\prime, k_1|p_2, s_2, c_2\rangle \Big],\ 
\label{m2-2}
\eea
where  the function $\mathcal{P}$ is related to $\tilde{\mathcal{P}}$ via, 
\be
\mathcal{P}(p_1,p_2) = \tilde{\mathcal{P}}(p_1,p_2) +\tilde{\mathcal{P}}(p_2,p_1). 
\ee
Note that while the amplitude $\tilde{\mathcal{P}}$ is not necessarily symmetric under the interchange of the two quarks, the function $\mathcal{P}$ is  symmetric by construction.  It  depends on longitudinal and transverse momentum of two quarks\footnote{With a mild abuse of notation we are using the same symbol $\mathcal{P}$ to denote the amplitude as a function of three-momenta as well as the function of transverse momenta and the longitudinal momentum fraction.}
\be
\mathcal{P}(p_1,p_2)\equiv \mathcal{P}\left(x_1  ,x_2, \p_1,\p_2\right), 
\ee
where $x_1$ and $x_2$ are the longitudinal light-cone fraction of the incoming quarks in the projectile nucleon wave function. The exact values of $x_1$ and $x_2$ are given later in \eq{xqs}.
Using \eq{m2-2}, we obtain, 
\bea 
&&|\langle \text{jet}(q), \text{jet}(q^\prime), \gamma(k_1),\gamma(k_2)| \text{two quarks} \rangle |^2= \frac{1}{4N^2_c} \sum_{\text{spin,color}} \int \int \int \int  \frac{d^3p_1}{(2\pi)^3}\frac{d^3p_2}{(2\pi)^3}
\frac{d^3p_1^\prime }{(2\pi)^3}\frac{d^3p_2^\prime}{(2\pi)^3}
 \mathcal{P}(p_1,p_2)\mathcal{P}^*(p^\prime_1,p^\prime_2)\nonumber\\
&\times&\Big[
\langle q, s, c; k_1|p_1,s_1, c_1\rangle \langle q^\prime, s^\prime, c^\prime, k_2|p_2, s_2, c_2\rangle 
 \langle p_2^\prime, s_2^\prime, c_2^\prime| q^\prime, s^\prime, c^\prime, k_2\rangle  \langle p_1^\prime,s_1^\prime, c_1^\prime |q, s, c; k_1\rangle \nonumber\\
&+&\langle q, s, c; k_1|p_1,s_1, c_1\rangle \langle q^\prime, s^\prime, c^\prime, k_2|p_2, s_2, c_2\rangle  
\langle p_2^\prime, s_2^\prime, c_2^\prime| q^\prime, s^\prime, c^\prime, k_1\rangle  \langle p_1^\prime,s_1^\prime, c_1^\prime |q, s, c; k_2\rangle \nonumber\\
&+&\langle q, s, c; k_2|p_1,s_1, c_1\rangle \langle q^\prime, s^\prime, c^\prime, k_1|p_2, s_2, c_2\rangle 
 \langle p_2^\prime, s_2^\prime, c_2^\prime| q^\prime, s^\prime, c^\prime, k_2\rangle  \langle p_1^\prime,s_1^\prime, c_1^\prime |q, s, c; k_1\rangle \nonumber\\
&+&\langle q, s, c; k_2|p_1,s_1, c_1\rangle \langle q^\prime, s^\prime, c^\prime, k_1|p_2, s_2, c_2\rangle 
 \langle p_2^\prime, s_2^\prime, c_2^\prime| q^\prime, s^\prime, c^\prime, k_1\rangle  \langle p_1^\prime,s_1^\prime, c_1^\prime |q, s, c; k_2\rangle 
\Big]. \
\label{cn}
\eea
 In the lowest order in the electro-magnetic $\alpha_{em}$ and the strong $\alpha_s$ coupling constants the $q\rightarrow q\gamma$ amplitude can be written in the following formal form, 
\bea
\langle q(q),\gamma(k_1)|q(p)\rangle
&=&-e_q\bar{u}(q)\Big[\mathcal{F}(q;p-k_1)G_F^0(p-k_1)\slashed{\epsilon}(k_1) +  \slashed{\epsilon}(k_1)G_F^0(q+k_1) \mathcal{F}(q+k_1,p)  \Big]u( p), \label{am1} \ 
\eea
where $e_q$ is the fractional electric charge of the projectile quark, $G_F^0$ is the free Feynman propagator of a quark with mass $m$. In the above $u$ and $\epsilon_\mu$ denote the quark free spinor and the photon polarization vector respectively. In the above, the operator matrix $\mathcal{F}$ contains the interaction between the quark and the colored glass condensate target, which resums multiple interactions with the background CGC field \cite{g-cgc0,g-cgc1}.  Assuming that the target is moving in the positive $z$ direction, we have \cite{pho-cgc1}, 
\be 
\mathcal{F}(q;p)=2\pi\delta(q^--p^-)\gamma^- sign(p^-)\int d^2\x \big[U(\x)-1\big]e^{i(\q-\p)\cdot\x}, \label{tf}
\ee 
where $U(\x)$ is a unitary matrix in fundamental representation of $SU(N_c)$ - the scattering matrix of a quark on the colored glass condensate target:
\be 
U(x)=T \exp\left(-ig^2\int dx^-\frac{1}{{\bf\nabla}^2}\rho_a(x^-,\x)t^a\right).
\ee 
Here $\rho$ is the density of the color sources in the target and $t^a$ is the generator of $SU(N_c)$ in the fundamental representation. 
Using the definition of $\mathcal{F}$ in \eq{tf}, one can rewrite the amplitude as, 
\bea
\langle q(q),\gamma(k_1)|q(p)\rangle &=&-ie_q\bar{u}({\bf q})\Big[\frac{\gamma^-
   (\s{p}-\s{k_1}+m)\s{\epsilon}(k_1)}
{(p-k_1)^2 -m^2}
+\frac{\s{\epsilon}(k_1)(\s{q}+\s{k_1}+m) \gamma^-}
{(q+k_1)^2-m^2}\Big]u({\bf p})\nonumber\\
&\times& 2\pi \delta(q^-+k^-_1-p^-)\int d^2{\x}\big[U({\x})-1\big]e^{i(\q+\k_1-\p_T)\cdot{\x}},\nonumber\\
&\approx&-ie_q\bar{u}({\bf q})
\gamma^-u({\bf p})\left[\frac{q\cdot\epsilon}{q\cdot k_1}-\frac{p\cdot\epsilon}{p\cdot k_1}\right]
2\pi \delta(q^-+k^-_1-p^-)\int d^2{\x}\big[U({\x})-1\big]e^{i(\q+\k_1-\p_T)\cdot{\x}}, \label{am-di}
\eea
where in the last line we employed the soft approximation, namely assuming that $|k_1|<|p-q|$.
In order to calculate the cross-section \eq{m1} we first  need to perform the color charge averaging of the expression \eq{cn} over the target  CGC field.  This is usually done using either the McLerran-Vengopalan model \cite{mv} defined by the weight function, 
\be 
W[\rho] =T \exp\left(-\int dx^-d^2{\bf x_T}\frac{\rho_a(x^-,\x)\rho^a(x^-,\x)}{2\mu^2(x^-)}\right),
\ee 
or using numerical solutions of Balitsky-Kovchegov equation \cite{bk}. For our present purposes the exact weight function does not matter. It is important though that any high energy/density averaging procedure does not affect the spin dependence in \eq{m1}.
\subsection{Tracing over color}
Using \eq{am-di} we can perform the summation over color indices and averaging over the target color field in \eq{cn}. A generic term in the integrand of \eq{cn} has the following structure,
\bea
\mathcal{I}&=&\Big{\langle} \sum_{\text{spin }} \sum_{\text{color}} \langle q, s, c; k_i|p_1,s_1, c_1\rangle \langle p^\prime_1, s^\prime_1, c^\prime_1|q, s, c; k_e\rangle 
\langle q^\prime, s^\prime, c^\prime; k_j|p_2, s_2, c_2\rangle  \langle p^\prime_2, s^\prime_2, c_2^\prime |  q^\prime, s^\prime, c^\prime; k_f\rangle \Big{\rangle_{\rho}}, \nonumber\\
&=& (-ie_q)^4 \delta_{tot}\times \mathcal{M}_{\text{spin}}\times \Big{\langle}\sum_{c_1,c_2,c_1^\prime,c_2^\prime} \left( \dots \right)_{c_1 c^\prime_1} \left( \dots \right)_{c_2 c^\prime_2} \Big{\rangle_{\rho}}
=(-ie_q)^4 \delta_{tot} \mathcal{M}_{\text{spin}} \frac{N_c^2}{N_c+1}\left(N_c N_F^{(2)}\times N_F^{(2)}+N_F^{(4)} \right), \label{t-p} \nonumber\\
\eea
where indices $(i,j=1,2)$ and $(e,f=1,2)$ denote the  two produced photons (note that $i\ne j$ and $e\ne f$) in the amplitude and conjugate  amplitude respectively, see \fig{f2}. The factor $\mathcal{M}_{\text{spin}}$ contains the spin summation (as defined in \eq{spin}) and $\delta_{tot}$ is given by, 
\be 
 \delta_{tot} = (2\pi)^4 \delta(q^-+k^-_i-p^{-}_1) \delta(q^-+k^-_e-p^{\prime -}_1)\delta(q^{\prime -}+k^-_j-p^{-}_2)\delta(q^{\prime -}+k^-_f-p^{\prime -}_2).  \label{e-delta}
\ee
In \eq{t-p}, $N_F^{(2)}$ and $N_F^{(4)}$  are  the traces of two (dipole) and four (quadrupole) light-like fundamental Wilson lines
in the background of the color fields of the target nucleus (or proton) respectively 
\bea
N_F^{(2)}(\b,\r_, x_g) &=& \frac{1}{N_c} \, \langle Tr [1 - U^{\dagger} (\x ) U (\y) ] \rangle_{x_g},  \nonumber\\
N_F^{(4)}(\b,\r,\b^\prime,\r^\prime, x_g) &=& \frac{1}{N_c} \, \langle Tr [1 - U^{\dagger} (\x) U (\y) U^{\dagger} (\x^\prime) U (\y^\prime)] \rangle_{x_g}, \label{d-q}\
\eea
where the vector ${\bf b}\equiv ({\bf x} + {\bf y})/2$ is the impact parameter of the dipole relative to the target and ${\bf r}\equiv {\bf x} - {\bf y}$ is the dipole transverse vector.
 Note that the expectation values on the right hand side are calculated over the ensemble of target fields evolved up to rapidity $y_g=\ln 1/x_g$.    The target in principle is evolved  by the JIMWLK \cite{jimwlk} or BK \cite{bk} equations.  The parameter $x_g$  can be related to the rapidities and transverse momenta of the prompt photons and final-state quarks via energy-momentum conservation \cite{kr}.  In the following for notational simplicity we drop the explicit label $x_g$ on the dipole and quadrupole amplitudes. 
 
The explicit expression for $\mathcal{I}$ in \eq{t-p} is given by, 
\bea
\mathcal{I}&=& (-ie_q)^4\delta_{tot} \mathcal{M}_{\text{spin}} \frac{N_c}{N_c+1}\Bigg[ \nonumber\\
&&\Big{\langle}\Tr [ \left(U(q+k_i-p_1)-1\right)\cdot\left(U^{\dagger}(p_1^\prime-q-k_e)-1\right) ]  \Big{\rangle_{\rho} }\Big{\langle}\Tr [ \left(U(q^\prime+k_j-p_2)-1\right)\cdot\left(U^{\dagger}(p_2^\prime-q^\prime-k_f)-1\right) ]  \Big{\rangle_{\rho}}  \nonumber\\
&+&\Big{\langle}\Tr \Big[ \left(U(q+k_i-p_1)-1\right)\cdot\left(U^{\dagger}(p_1^\prime-q-k_e)-1\right)\cdot 
\left(U(q^\prime+k_j-p_2)-1\right)\cdot \left(U^{\dagger}(p_2^\prime-q^\prime-k_f)-1\right) \Big] \Big{\rangle_{\rho}}\Bigg].    \label{dq}\
\eea
The first and the second terms in \eq{dq} correspond to the Fourier transformed dipole and quadrupole scattering amplitudes defined in \eq{d-q}, respectively. It is useful to rewrite  the above expression in terms of the dipole transverse separation vector $\r$ (and $\r^\prime$) and the impact parameter $\b$ (and $\b^\prime$), 
\bea
&&\frac{\mathcal{I}\times (N_c+1)}{e_q^4 \delta_{tot} \mathcal{M}_{\text{spin}}N_c }=  \int d^2\r\, d^2\b\, e^{i\r. \left(\q+\frac{1}{2}\left(\k_i+\k_e-\p_1-\p_1^\prime\right)\right)}\, 
e^{i\b. \left(\k_i-\k_e+\p_1^\prime-\p_1\right)} \Big{\langle}\Tr [ \left(U(\b+\r/2)-1\right)\cdot\left(U^{\dagger}(\b-\r/2)-1\right) ]  \Big{\rangle_{\rho} }\nonumber\\
&&\times\int d^2\r^\prime \, d^2\b^\prime\, e^{i\r^\prime. \left(\q^\prime+\frac{1}{2}\left(\k_j+\k_f-\p_2-\p_2^\prime\right)\right)}\, 
e^{i\b^\prime. \left(\k_j-\k_f+\p_2^\prime-\p_2\right)} \Big{\langle}\Tr [ \left(U(\b^\prime+\r^\prime/2)-1\right)\cdot\left(U^{\dagger}(\b^\prime-\r^\prime/2)-1\right) ]  \Big{\rangle_{\rho} }\nonumber\\
&&+\int d^2\r\, d^2\b\, d^2\r^\prime \, d^2\b^\prime\,    e^{i\r. \left(\q+\frac{1}{2}\left(\k_i+\k_e-\p_1-\p_1^\prime\right)\right)}\,
e^{i\r^\prime. \left(\q^\prime+\frac{1}{2}\left(\k_j+\k_f-\p_2-\p_2^\prime\right)\right)}\,  
e^{i\b. \left(\k_i-\k_e+\p_1^\prime-\p_1\right)}\, e^{i\b^\prime. \left(\k_j-\k_f+\p_2^\prime-\p_2\right)}\nonumber\\
&&\times \Big{\langle}\Tr \Big[ \left(U( \b+\r/2)-1\right)\cdot\left(U^{\dagger}(\b-\r/2)-1\right)\cdot 
 \left(U(\b^\prime+\r^\prime/2)-1\right)\cdot \left(U^{\dagger}(\b^\prime-\r^\prime/2 )-1\right) \Big] \Big{\rangle_{\rho}}.  \label{rb}\
 \eea
At large $N_c$, the term containing the quadrupole amplitude is suppressed relative to the one containing dipoles by a factor $1/N_c$. At leading order in $1/N_c$  one may therefore  ignore the quadrupole contribution.

Another simplification arises if we assume that the target is  uniform in the impact parameter space. This approximation may be appropriate for p-A scatterings, and is almost always employed in CGC based calculations. Under this assumption one can ignore the $b$ dependence in $N_F^{(2)}$, and  the integrals over $\b$ and $\b^\prime$ in \eq{rb} lead to delta functions. Therefore the cross-section has the structure, 
\be 
\mathcal{I} \propto\delta^2(\k_i-\k_e-{\mathbf\Delta_{1}}) \delta^2(\k_j-\k_f-{\mathbf\Delta_{2}}),  \label{del}
\ee 
where we have defined 
\bea 
{\mathbf\Delta_{1}} &=& \p_1-\p_1^\prime,  \nonumber\\
 {\mathbf\Delta_2 }&=& \p_2-\p_2^\prime, \label{del-def}\
\eea 
with ($\p_1$, $\p_2$)  and  ($\p_1^\prime $, $\p_2^\prime$)  being the transverse momenta of two projectile quarks in the amplitude and its conjugate amplitude, respectively.  
The momentum ${\mathbf\Delta_{1,2}}$ is the difference of the momenta of two partons from the wave function of the colliding hadron in the amplitude and the amplitude conjugated. Note that the difference of parton transverse momenta within the parton pair is not conserved. Thus at large $N_c$ we have, 
\begin{equation}
\mathcal{I}=(-ie_q)^4 (2\pi)^4  N^2_c \delta_{tot} \mathcal{M}_{\text{spin}}\delta^2(\k_i-\k_e-{\mathbf\Delta_{1}}) \delta^2(\k_j-\k_f-{\mathbf\Delta_{2}})N_F^{(2)} \left(\q+\k_i-\p_1\right)N_F^{(2)} \left(\q^\prime+\k_j-\p_2\right).  \label{delta2}
\end{equation}

\subsection{Tracing over spin}
Now we turn to the spin summation in the expression \eq{cn}. The matrix element $\mathcal{M}_{\text{spin}}$ in \eq{t-p} is given by,   
 \bea
\mathcal{M}_{\text{spin}} &=& \sum_{\alpha,\beta,\gamma,\eta  }\,\, \sum_{ s_1, s_2^\prime, s_2,s_2^\prime} \bar{u}^{s_1^\prime}_{\alpha}(p_1^\prime)  \mathcal{A}_{\alpha \beta} (q, k_i, k_e, p_1, p_1^\prime) u^{s_1}_{\beta}(p_1)\bar{u}^{s_2^\prime}_{\gamma}(p_2^\prime)  \mathcal{A}_{\gamma \eta} (q^\prime, k_j, k_f, p_2, p_2^\prime)  u^{s_2}_{\eta}(p_2),  \label{spin}\   
\eea
where the matrix functions $\mathcal{A}$ is defined as follows, 
\bea
\mathcal{A} (q, k_i, k_e, p, p^\prime)&=&
\gamma^- (\s{q}+m)\gamma^{-}\left[\frac{q\cdot\epsilon_i}{q\cdot k_i}-\frac{p\cdot\epsilon_i}{p\cdot k_i}\right]\left[\frac{q\cdot\epsilon^*_e}{q\cdot k_e}-\frac{p'\cdot\epsilon_e^*}{p'\cdot k_e}\right]. \label{spin-a}
\eea
The summation over photon polarization is implicit in \eq{spin}. The mass-term in the spin matrix element in the above equation is inherited  from the quark-propagator in \eq{am-di}.  However, in the high-energy limit employed here, the mass term in fact is irrelevant, since $(\gamma^-)^2=0$, and it thus disappears from all the following formulae. All the terms in \eq{cn} have similar structure to \eq{t-p} and can be written out explicitly using  Eqs.\,(\ref{e-delta},\ref{rb},\ref{delta2},\ref{spin}). Therefore, at large $N_c$  we obtain, 
\bea
&&|\langle \text{jet}(q), \text{jet}(q^\prime), \gamma(k_1),\gamma(k_2)| \text{two quarks} \rangle |^2= 
e_q^4 \pi^2\sum_{\text{spin}}\int \int  \frac{d^3 p_1}{(2\pi)^3}\frac{d^3 p_2}{(2\pi)^3}N^{(2)} \left(\q+\k_1-\p_1\right)N^{(2)} \left(\q^\prime+\k_2-\p_2\right)  \nonumber\\
&\times&\delta(p_1^--q^--k_1^-)\delta(p_2^--q^{\prime -}-k_2^-) \Bigg\{\mathcal{P}(x_1,x_2,\p_1,\p_2)\mathcal{P}^*(x_1,x_2, \p_1,\p_2)     \nonumber\\
&\times&\bar{u}(p_1)\gamma^- \s{q}\gamma^-u( p_1)\bar{u}(p_2)\gamma^- \s{q'}\gamma^-u( p_2)\left|\frac{q\cdot\epsilon_1}{q\cdot k_1}-\frac{p_1\cdot\epsilon_1}{p_1\cdot k_1}\right|^2
\left|\frac{q'\cdot\epsilon_2}{q'\cdot k_2}-\frac{p_2\cdot\epsilon_2}{p_2\cdot k_2}\right|^2\nonumber\\
&+& \mathcal{P}(x_1,x_2,\p_1,\p_2)\mathcal{P}^*(x_1^\prime,x_2^\prime, \p_1+\k_2-\k_1,\p_2+\k_1-\k_2)\nonumber\\
&\times&\bar{u}(p_1+k_2-k_1)\gamma^- \s{q}\gamma^-u( p_1)\bar{u}(p_2+k_1-k_2)\gamma^- \s{q'}\gamma^-u( p_2)\nonumber\\
&\times&\left[\frac{q\cdot\epsilon_1}{q\cdot k_1}-\frac{p_1\cdot\epsilon_1}{p_1\cdot k_1}\right]\left[\frac{q\cdot\epsilon^*_2}{q\cdot k_2}-\frac{     (p_1+k_2-k_1)\cdot\epsilon_2^*}{(p_1+k_2-k_1)\cdot k_2}\right]\left[\frac{q'\cdot\epsilon_2}{q'\cdot k_2}-\frac{p_2\cdot\epsilon_2}{p_2\cdot k_2}\right]\left[\frac{q'\cdot\epsilon^*_1}{q^*\cdot k_1}-\frac{(p_2+k_1-k_2)\cdot\epsilon_1^*}{(p_2+k_1-k_2)\cdot k_1}\right]\Bigg\}\nonumber\\
 &+&\left(k_1\leftrightarrow k_2\right) , \label{final2}\
\eea
where we performed the integrals over $p_1^\prime$ and $p_2^\prime$ in  \eq{cn} using the delta functions in Eqs.\,(\ref{e-delta},\ref{delta2}). In the above, the light-cone parameters $x_1,x_2, x_1^\prime, x_2^\prime$ in ${\cal P}$ and ${\cal P^*}$ are given by, 
\bea \label{xqs}
x_1&=&\frac{q^- + k_1^-}{\sqrt{s/2}},\hspace{3.5cm}   x_2=\frac{q^{\prime -} + k_2^-}{\sqrt{s/2}},\nonumber\\
x_1^\prime&=&\frac{q^- + k_2^-}{\sqrt{s/2}}=x_1+\frac{k_2^--k_1^-}{\sqrt{s/2}},\hspace{1cm}   x_2^\prime=\frac{q^{\prime -} + k_1^-}{\sqrt{s/2}}=x_2+\frac{k_1^--k_2^-}{\sqrt{s/2}}, \
\eea
 with $\sqrt{s}$ being the nucleon-nucleon
center-of-mass energy, and we have
\be
x_1+x_2 \le 1. 
\ee
We now further simplify the expression in \eq{final2}. First, we commute one of the $\gamma^-$ through $\s{q}$ and use the fact that $\gamma^-\gamma^-=0$. We also perform the sum over the photon polarization using the completeness  of the photon polarization vectors, and neglecting the quark mass. Therefore, we obtain, 
\bea
&&|\langle \text{jet}(q), \text{jet}(q^\prime), \gamma(k_1),\gamma(k_2)| \text{two quarks} \rangle |^2=  e_q^4 16\pi^2\sum_{\text{spin}}\int \int  \frac{d^3 p_1}{(2\pi)^3}\frac{d^3 p_2}{(2\pi)^3}N^{(2)} \left(\q+\k_1-\p_1\right)N^{(2)} \left(\q^\prime+\k_2-\p_2\right)  \nonumber\\
&\times&\delta(p_1^--q^--k_1^-)\delta(p_2^--q^{\prime -}-k_2^-)q^-q^{\prime -}
\Bigg\{\mathcal{P}(x_1,x_2,\p_1,\p_2)\mathcal{P}^*(x_1,x_2, \p_1,\p_2)  \bar{u}(p_1)\gamma^-u( p_1)\bar{u}(p_2)\gamma^-u( p_2) \nonumber\\
&\times& \frac{q\cdot p_1}{q\cdot k_1 p_1\cdot k_1} \frac{q'\cdot p_2}{q'\cdot k_2 p_2\cdot k_2}
\nonumber\\
&+& \mathcal{P}(x_1,x_2,\p_1,\p_2)\mathcal{P}^*(x_1^\prime,x_2^\prime, \p_1+\k_2-\k_1,\p_2+\k_1-\k_2) \bar{u}(p_1+k_2-k_1)\gamma^-u( p_1)\bar{u}(p_2+k_1-k_2)\gamma^-u( p_2)\nonumber\\
&\times&\left[\frac{q\cdot q'}{q\cdot k_1 q'\cdot k_1}+\frac{p_1\cdot(p_2+k_1-k_2)}{p_1\cdot k_1 (p_2+k_1-k_2)\cdot k_1}-
\frac{q\cdot(p_2+k_1-k_2)}{q\cdot k_1 (p_2+k_1-k_2)\cdot k_1 }-\frac{p_1\cdot q'}{p_1\cdot k_1 q'\cdot k_1}
\right]\nonumber\\
&\times&\left[\frac{q\cdot q'}{q\cdot k_2 q'\cdot k_2}+\frac{p_2\cdot(p_1+k_2-k_1)}{p_2\cdot k_2 (p_1+k_2-k_1)\cdot k_2 }-
\frac{q'\cdot(p_1+k_2-k_1)}{q'\cdot k_2 (p_1+k_2-k_1)\cdot k_2 }-\frac{p_2\cdot q}{p_2\cdot k_2 q\cdot k_2}
\right]  \Bigg\}\nonumber\\
 &+&\left(k_1\leftrightarrow k_2\right) , \label{final1}\
\eea
Note that in the above expression each Dirac spinor carries an index $s$ and these indices are summed over completely independently. However, the spin structure can be simplified further in the high energy limit. At high energy the incoming quark, outgoing quark and photon are practically collinear, since the scattering angle of the quark is very small at finite transverse momentum transfer. In such kinematics angular momentum conservation requires that  the helicities of the incoming and outgoing quarks are opposite.
This in turns  means that $s_1=s_1^\prime$ and $s_2=s_2^\prime$, see \fig{f2}. Using the high-energy properties of Dirac spinors, one can perform the remaining spin summation in \eq{final1} and obtain, 
\bea
&&|\langle \text{jet}(q), \text{jet}(q^\prime), \gamma(k_1),\gamma(k_2)| \text{wo quarks} \rangle |^2 = e_q^4 (16\pi)^2\int \int  \frac{d^3 p_1}{(2\pi)^3}\frac{d^3 p_2}{(2\pi)^3}N^{(2)} \left(\q+\k_1-\p_1\right)N^{(2)} \left(\q^\prime+\k_2-\p_2\right)  \nonumber\\
&\times&\delta(p_1^--q^--k_1^-)\delta(p_2^--q^{\prime -}-k_2^-)q^-q^{\prime -}
\Bigg\{\mathcal{P}(x_1,x_2,\p_1,\p_2)\mathcal{P}^*(x_1,x_2, \p_1,\p_2) p_1^-p_2^-  \frac{q\cdot p_1}{q\cdot k_1 p_1\cdot k_1} \frac{q'\cdot p_2}{q'\cdot k_2 p_2\cdot k_2}
\nonumber\\
&+& \mathcal{P}(x_1,x_2,\p_1,\p_2)\mathcal{P}^*(x_1^\prime,x_2^\prime, \p_1+\k_2-\k_1,\p_2+\k_1-\k_2)\sqrt{p_1^-(p_1^-+k_2^--k_1^-)p_2^-(p_2^-+k_1^--k_2^-)}\nonumber\\
&\times&\left[\frac{q\cdot q'}{q\cdot k_1 q'\cdot k_1}+\frac{p_1\cdot(p_2+k_1-k_2)}{p_1\cdot k_1 (p_2+k_1-k_2)\cdot k_1 }-
\frac{q\cdot(p_2+k_1-k_2)}{q\cdot k_1 (p_2+k_1-k_2)\cdot k_1 }-\frac{p_1\cdot q'}{p_1\cdot k_1 q'\cdot k_1}
\right]\nonumber\\
&\times&\left[\frac{q\cdot q'}{q\cdot k_2 q'\cdot k_2}+\frac{p_2\cdot(p_1+k_2-k_1)}{p_2\cdot k_2 (p_1+k_2-k_1)\cdot k_2 }-
\frac{q'\cdot(p_1+k_2-k_1)}{q'\cdot k_2 (p_1+k_2-k_1)\cdot k_2 }-\frac{p_2\cdot q}{p_2\cdot k_2 q\cdot k_2}
\right]\Bigg\}  \nonumber\\
 &+&\left(k_1\leftrightarrow k_2\right).  \label{final}
\eea
The products  $\mathcal{P}\mathcal{P}^*$ that appear  in our final expression \eq{final} can be interpreted in terms of generalized double transverse momentum-dependent parton distributions (2GTMD) of the projectile hadron (denoted by $\mathcal{T}$, see \eq{tdef}). 
First off note, that if we discard the simplifying assumption about the factorizability of the reduced density matrix Eq. (\ref{product}), our final formulae would remain the same apart of the substitution
\be\label{dens}
 \mathcal{P}\left(x_1,x_2,\p_1,\p_2\right)\mathcal{P}^*\left(x_1^\prime,x_2^\prime, \p^\prime_1,\p^\prime_2\right)\rightarrow {\rm tr}\left[\tilde{\mathcal{R}}\left(x_1,x_2,\p_1,\p_2,x_1^\prime,x_2^\prime, \p^\prime_1,\p^\prime_2\right)\right],\ee
 where on the right hand side the density matrix is traced over the spin and color.
 
 Our approximation of the translational invariance of the  nuclear wave function  in the impact parameter space (which led to the delta functions in \eq{del}) means that the total transverse momenta carried by quarks in the amplitude and complex conjugate amplitude are equal. As a result 
the first term in \eq{final}, we have ${\mathbf \Delta_1}={\mathbf \Delta_2}=0$  while  in the second term (the correlated part), we have ${\mathbf \Delta_1}={\mathbf \Delta_2}={\mathbf \Delta}\ne0$, where ${\mathbf \Delta}\equiv \k_2-\k_1$.
Additionally note that in the soft approximation which we are employing throughout, $x^\prime_1\approx x_1; \ x^\prime_2\approx x_2$.
Thus the basic quantity that appears in Eq. (\ref{final1}) is
\bea\label{tdef}
\mathcal{T}(x_1,x_2, \p_1,\p_2, \Delta ) &\equiv&  {\rm tr}\left[\tilde{\mathcal{R}}\left(x_1,x_2,\p_1,\p_2,x_1,x_2, \p_1+{\mathbf \Delta},\p_2-{\mathbf \Delta}\right)\right],\nonumber\\
&=&
\mathcal{P}\left(x_1,x_2,\p_1,\p_2\right)\mathcal{P}^*(x_1,x_2, \p_1+{\mathbf \Delta},\p_2-{\mathbf \Delta}).  
\eea
In terms of the wave function of the hadron it is defined as (suppressing spin and color indices)
 \bea \label{2dpg}
\mathcal{T}(x_1,x_2, \p_1,\p_2, \Delta)&=& 
\sum_{n=3}^{\infty} 
\int {\displaystyle \prod_{i\ne 1,2}}  \frac{d^2 \p_i}{(2\pi)^2} \int_0^1 {\displaystyle \prod_{i\ne 1,2} }dx_i 
\nonumber\\
&\times&  \Psi_n\left(x_1, x_2, \dots, \p_1,\p_2, \dots \right) \Psi_n^{+}\left(x_1, x_2, \dots, \p_1+\Delta, \p_2-\Delta, \dots \right)   \nonumber\\
&\times& (2\pi)^3 \delta\left(\sum_{i=1}^{n} x_i-1\right)\delta\left(\sum_{i=1}^{n} \p_i \right),\ 
 \eea
where $\Psi_n$ is the normalized n-parton wave-function.  
  It is related to  the  generalized double parton distribution (2GPD) \cite{marke}  in a simple way
\bea D(x_1,x_2, \mu_1^2,\mu_2^2, \Delta)&=& 
 \int  \frac{d^2 \p_1}{(2\pi)^2}\frac{d^2 \p_2}{(2\pi)^2} \theta(\mu_1^2-\p_1^2) \theta(\mu_2^2-\p_2^2)  \mathcal{T} (x_1,x_2, \p_1,\p_2, \Delta), 
 \eea
 where $\mu_1^2$ and $\mu_2^2$ are the virtualities of the two quarks.
This nonperturbative object, 2GPD denoted by $D$,  appears in the calculations involving DPS in pQCD in the collinear factorization framework, for example  the four-jet production in proton-proton collisions \cite{marke}, see also Refs.\,\cite{2gpd-1,2gpd-2,2gpd-3,2gpd-4}.

Using the above definitions and  Eqs.\,(\ref{m1},\ref{final}), we can re-write the cross-section of double photon-quark pair production in the following general form, 
\bea  
d\sigma^{qq+A\to\gamma\gamma+qq}&=&
 \int \int  \frac{d^3 p_1}{(2\pi)^3}\frac{d^3 p_2}{(2\pi)^3} \Bigg[\mathcal{T}\left(x_1,x_2, \p_1,\p_2, 0\right)\, d\sigma^{q(p_1)+A\to\gamma(k_1)+q(q)}\times d\sigma^{q(p_2)+A\to\gamma(k_2)+q(q^\prime)}
\nonumber\\
&+& \mathcal{T}\left(x_1,x_2, \p_1,\p_2, \Delta \right)\, d\sigma^{\text{Interference}}\Bigg], \label{final-g}\
 \eea
where $d\sigma^{q+A\to\gamma+q}$ is the cross-section of the single prompt photon-quark production in q+A collisions calculable via diagrams in \fig{f0} and can be immediately extracted from our final expression in \eq{final}. The cross-section of $d\sigma^{q+A\to\gamma+q}$ obtained here is consistent with the soft approximation in Ref.\,\cite{kr}. 
 In the approximation of uncorrelated partons, we have 
\be \label{dps1}
\mathcal{T}\left( x_1, x_2, \p_1, \p_2,\Delta \right)\approx G_{GTMD}\left( x_1,  \p_1,  \Delta \right) G_{GTMD}\left( x_2,  \p_2 , \Delta\right),  
\ee
where  $G_{GTMD}\left( x_1,  \p_1, \Delta\right)$ is the one-particle generalized transverse
momentum-dependent parton distribution (GTMD) \cite{gtmd1,gtmd2,gtmd3,gtmd4,gtmd5,gtmd6}. Under the assumption of uncorrelated partons in the projectile hadron, the first term in Eqs.\,(\ref{final},\ref{final-g}), can be factorized into two independent cross-sections for prompt photon-quark production. Hence, the first part in \eq{final}  (and in \eq{final-g}) contains  the contributions of independent production. 
 The second term in \eq{final} (and in \eq{final-g}) cannot be factorized into two independent terms  even if the partons in the projectile wave function are uncorrelated. This term leads to nontrivial correlations between two produced photons, whose nature we discuss in the next section. Note that the correlated part (second term) corresponds to the interference diagrams where the produced photons in the amplitude and its conjugated  amplitude have different momenta, see \fig{f2}.  It is remarkable that the correlations of two produced photons are explicitly related to the fact that  $\Delta\ne 0$ in the second term.

\section{Hanbury Brown and Twiss (HBT) correlations in inclusive di-photon cross-section}
 The inclusive di-photon cross-section is obtained by substituting the expression in \eq{final} into \eq{m1}, and integrating over $q$ and  $q^\prime$. Thus we obtain,
\bea
&&\frac{d\sigma^{qq+A\to\gamma\gamma+X}}{d^3k_1 d^3 k_2}= 
\frac{e_q^4}{ (2\pi)^6} \frac{1}{k_1^- k_2^-} \int \int \int \int \frac{d^3 p_1}{(2\pi)^3}\frac{d^3 p_2}{(2\pi)^3} 
 \frac{d^2 \q}{(2\pi)^2}\frac{d^2 \q^\prime}{(2\pi)^2} 
N^{(2)} \left(\q+\k_1-\p_1\right)N^{(2)} \left(\q^\prime+\k_2-\p_2\right)  \nonumber\\
\nonumber\\
&&\Bigg\{\mathcal{T}\left(x_1,x_2,\p_1,\p_2, 0\right)\,  \frac{q\cdot p_1}{q\cdot k_1 p_1\cdot k_1} \frac{q'\cdot p_2}{q'\cdot k_2 p_2\cdot k_2}
\nonumber\\
&+& \mathcal{T}\left(x_1,x_2, \p_1,\p_2, \Delta \right)\frac{\sqrt{p_1^-(p_1^-+\Delta^-)p_2^-(p_2^- -\Delta^-)}}{ p_1^-p_2^-}\nonumber\\
&\times&\left[\frac{q\cdot q'}{q\cdot k_1 q'\cdot k_1}+\frac{p_1\cdot(p_2-\Delta)}{p_1\cdot k_1 (p_2-\Delta)\cdot k_1}-
\frac{q\cdot(p_2-\Delta)}{q\cdot k_1 (p_2-\Delta)\cdot k_1 }-\frac{p_1\cdot q'}{p_1\cdot k_1 q'\cdot k_1}
\right]\nonumber\\
&\times&\left[\frac{q\cdot q'}{q\cdot k_2 q'\cdot k_2}+\frac{p_2\cdot(p_1+\Delta)}{p_2\cdot k_2 (p_1+\Delta)\cdot k_2 }-
\frac{q'\cdot(p_1+\Delta)}{q'\cdot k_2 (p_1+\Delta)\cdot k_2 }-\frac{p_2\cdot q}{p_2\cdot k_2 q\cdot k_2}
\right]\Bigg\},  \label{inclusive}
\eea
where 
\be q^-\equiv p_1^--k_1^-;\ \ \ \ q^{\prime-}\equiv p_2^--k_2^-;\ \ \ \  \Delta=k_2-k_1.
\ee
The second term in this expression, the interference term expresses the HBT correlations. We note that under the naive parton model  assumption that the transverse momenta of all incoming partons vanish, we have $\mathcal{T}\left(x_1,x_2, \p_1,\p_2, \Delta \right)\propto \delta(\Delta)$ and thus the HBT peak has zero width.

To understand the qualitative features of this expression beyond this naive approximation we consider the following special  kinematics. First we take the two photons to be soft  , $k^-_{1(2)}\ll p^-_1,p^-_2$. We also assume that the transverse momentum of the two photons are large, but are not too different from each other  $|\k_1+\k_2|\gg|\k_1-\k_2|$.   In the spirit of parton model that the intrinsic transverse momentum in the proton wave function is small, and thus the integration over $\p_1$ and $\p_2$ is dominated by the region $|\p_1|,|\p_2|\ll |\k_1|\approx|\k_2|$. Additionally we assume that the dipole scattering amplitude is saturated, and thus the momentum transfer is strongly peaked at $Q_s\ll k$. In the following we use the notation $k=|\k_1+\k_2|/2$. In this kinematics  we obtain, 
\begin{equation}
\left[\frac{q\cdot q'}{q\cdot k_1 q'\cdot k_1}+\frac{p_1\cdot(p_2-\Delta)}{p_1\cdot k_1 (p_2-\Delta)\cdot k_1}-
\frac{q\cdot(p_2-\Delta)}{q\cdot k_1 (p_2-\Delta)\cdot k_1 }-\frac{p_1\cdot q'}{p_1\cdot k_1 q'\cdot k_1}
\right]\approx-\frac {4(k^-_1)^2}{s/2}\frac{1}{x_1x_2}\frac{\q'\cdot\q}{k^4}.
\end{equation}
\begin{equation}
\left[\frac{q\cdot q'}{q\cdot k_2 q'\cdot k_2}+\frac{p_2\cdot(p_1+\Delta)}{p_2\cdot k_2 (p_1+\Delta)\cdot k_2 }-
\frac{q'\cdot(p_1+\Delta)}{q'\cdot k_2 (p_1+\Delta)\cdot k_2 }-\frac{p_2\cdot q}{p_2\cdot k_2 q\cdot k_2}
\right]\approx-\frac {4(k^-_2)^2}{s/2}\frac{1}{x_1x_2}\frac{\q'\cdot\q}{k^4}.
\end{equation}
For the interference contribution to the cross-section we obtain
\bea
&&\frac{d\sigma^{qq+A\to\gamma\gamma+X}}{d^3k_1 d^3 k_2}|_{\text{interference}} \approx
\frac{e_q^4}{ (2\pi)^6} k_1^- k_2^-\frac{16}{(2\pi)^2k^8}\nonumber\\
&\times& \int \frac{d^2 \p_1}{(2\pi)^2}\frac{d^2 \p_2}{(2\pi)^2} \frac{dx_1}{x_1}\frac{dx_2}{x_2}
 \frac{d^2 \q}{(2\pi)^2}\frac{d^2 \q^\prime}{(2\pi)^2} \frac{(\q'\cdot\q)^2}{ p_1^-p_2^-}
N^{(2)} \left(\q+\k_1-\p_1\right)N^{(2)} \left(\q^\prime+\k_2-\p_2\right)  \mathcal{T}\left(x_1,x_2, \p_1,\p_2, \Delta \right).\
\eea
Assuming rotational invariance of the dipole scattering amplitude we can estimate the average value of momentum as
\begin{equation}
 \int\frac{d^2 \q}{(2\pi)^2}\q_i\q_jN^{(2)} \left(\q+\k_1-\p_1\right)= \int\frac{d^2 \q}{(2\pi)^2}(\q-\k_1+\p_1)_i(\q-k_1+p_1)_jN^{(2)} \left(\q\right)=\frac{1}{2}\delta_{ij}\int\frac{d^2 \q}{(2\pi)^2}\q^2N^{(2)} \left(\q\right)=\frac{1}{2}\delta_{ij}Q_s^2S_{eff},
 \end{equation}
where $S_{eff}$ is an effective interaction area, $Q_s$ is saturation scale of the system and we have used the fact that 
\begin{equation}\int\frac{d^2 \q}{(2\pi)^2}N^{(2)} \left(\q\right)=0, \ \ \ \ \int\frac{d^2 \q}{(2\pi)^2}\q_jN^{(2)} \left(\q\right)=0.
\end{equation}
Therefore we obtain, 
\bea \label{finalf}
\frac{d\sigma^{qq+A\to\gamma\gamma+X}}{d^3k_1 d^3 k_2}|_{\text{interference}} \approx
\frac{2e_q^4}{ (2\pi)^6} \frac{k_1^- k_2^-}{s}\frac{16Q_s^4S_{eff}^2}{(2\pi)^2k^8} \int \frac{d^2 \p_1}{(2\pi)^2}\frac{d^2 \p_2}{(2\pi)^2}\,dx_1\,dx_2\, \mathcal{T}\left(x_1,x_2, \p_1,\p_2, \Delta \right).
\eea
We did not indicate the virtuality of the double parton distribution in the above, but it is clearly given by the large momentum scale in the problem, which is the transverse momentum of the individual photons $k_T$. 

The form of the 2GTMD in \eq{finalf} is not known experimentally.
Nevertheless, the physics of the correlation present in \eq{finalf} is clearly that  of the Hanbury Brown and Twiss (HBT) effect. We find correlation between the  bosons (photons) emitted from uncorrelated  sources (quarks).  Indeed the behavior of this interference term is precisely  a typical HBT behavior. 
 The easiest way to see this is in the approximation where the two incoming partons are taken to be uncorrelated in the proton wave function in high energy limit. In this case we assume 
 \bea {\cal P}(x_1,x_2,\p_1,\p_2)={\cal P}(x_1,\p_1){\cal P}(x_2,\p_2).\eea
The exact shape of  $ {\cal P}(x,\p)$ does not matter much. The only important aspect of it,  is that it should reflect the existence of the nonperturbative distance scale $R$. This scale determines the physical size of the quark cloud in the proton and is thus naturally associated with the proton radius. In momentum space this means that the TMD should decrease beyond $p\sim R^{-1}$.
For illustrative purposes here we assume a simple Gaussian distribution for the intrinsic momentum dependence\footnote{Such a Gaussian distribution is supported by various phenomenological studies (see  for example Refs.\,\cite{gauss1,gauss2}). In this paper however we are using it merely as an illustration. }
\bea\label{pgaus}
{\cal P}(x,\p)\propto e^{-\frac{1}{2}R^2|\p|^2},
\eea
We then have
\bea\label{formfa}
\int \frac{d^2 \p_1}{(2\pi)^2}\frac{d^2 \p_2}{(2\pi)^2}\mathcal{T}\left(x_1,x_2, \p_1,\p_2, \Delta \right)=f_q(x_1)f_q(x_2)e^{-\frac{1}{2}R^2\Delta^2}, 
\eea
where the function $f_q(x_1)$ is the usual quark pdf (the virtuality is implicit). 
The interference thus leads to enhancement of the cross-section for $|\k_1-\k_2|<1/R$ - a typical HBT correlation behavior. The nonperturbative scale $R$ can be therefore directly measured by measuring photon correlations. The magnitude of the effect drops pretty fast at large transverse momentum of the photons, but presumably at $k_T\sim Q_s$ the interference piece  should not be  significantly  suppressed relative to the independent production piece.

Another popular assumption in the literature is to approximate the GPD by 
 $G_{GPD}\left( x_1,  \p^2_1, \Delta\right)\approx G\left(x_1,  \p^2_1\right) \mathcal{F}(\Delta)$ where $G$ is the conventional parton (quark) distribution of the nucleon  and $\mathcal{F}(\Delta)$  is the nonperturbative proton form-factor \cite{marke,mark2}. If this factorization is assumed to hold at any virtuality $p_1^2$, it is equivalent to 
a similar factorizable approximation for GTMD:
 $G_{GTMD}\left( x_1,  \p_1, \Delta\right)\approx G\left(x_1,  \p_1\right) \mathcal{F}(\Delta)$ where  $G\left(x_1,  \p_1\right)$ is the  transverse momentum dependent distribution (TMD).
This, via the use of \eq{dps1} again leads to \eq{formfa} with the Gaussian factor replaced by $\mathcal{F}(\Delta)$. The form factor is maximal at $\Delta=0$, and decreases on the momentum scale $\mu$, which  has the same physical meaning as the scale $R^{-1}$ introduced in \eq{pgaus}.

We note that a similar form factor for gluons was discussed in the literature  and the functional form
 $\mathcal{F}_g(\Delta)=\frac{1}{(\Delta^2/\mu^2+1)^2}$ was extracted from exclusive vector meson production with $\mu^2\approx 1\,\text{GeV}^2$ \cite{marke,mark2}. The value of the proton size extracted from $\mathcal {F}_g$ is rather small, which is consistent with many other experimental indications of a small gluonic radius of the proton \cite{boris}. For di-photon HBT we expect a different, and larger transverse distance scale to dominate the HBT correlations. 
 
 \section{Conclusions}
 In this paper we have developed the hybrid calculational approach to forward particle production to include DPS processes in the saturated environment. The main technical ingredient that appears in this approach is the generalized double transverse momentum-dependent parton distribution (2GTMD) function. In the ``Hybrid'' approach, the DPS means two partons from the projectile hadron coherently colliding with the CGC shock wave.   Thus on the target side all multiple scattering interactions are resummed  in our calculation. In this sense we do not distinguish between interactions of a single, double or higher number of target partons. For that reason the nuclear 2GTMD (or 2GPD) does not appear as a distinct object in our calculation, and only  the 2GTMD (or 2GPD) of the projectile proton is relevant.  This  is  in contrast to study of  Ref.\,\cite{g1} where because the target was considered in the standard pQCD approach,  the nuclear 2GPDs contribution is separately identified and can be studied in multiple-jet production in pA collisions.  
 
 We studied in detail the di-photon correlations that arise due to the DPS process. We found that these correlations reflect the Hanbury Brown and Twiss effect, and lead to enhanced double photon production when the transverse momentum of the two photons are within the inverse proton radius of each other. At high momentum of produced photons the correlated piece decreases quite fast (as $1/k^8$), but it should give a significant enhancement when the photon momenta are not much larger than the target saturation scale.  We showed that the width of the HBT peak probes the transverse distance between the parton of the pair in the 2GTMDs. Therefore, the HBT measurements in two-particle production such as prompt photon pairs provide useful information about the nonperturbative 2GTMDs.

It would be interesting to compare the magnitude of the correlated cross-section we find here with the correlations generated through SPS \cite{kr}.  We did not attempt a quantitative comparison, since the 2GTMD's are not known with significant accuracy. It is interesting to note however, that parametrically the DPS contribution can be competitive with the SPS one, especially for intermediate transverse momentum photons with $|\k|$ not much larger than $Q_s$.  Although one requires two quarks to scatter, in the saturated regime where the quark scattering amplitude is of the order one, this is not suppressed by powers of $\alpha_s$. Additionally, the DPS is enhanced by a factor roughly equal to the number of quarks in the proton. Thus all in all the DPS contribution can be comparable to the SPS one.
 It would be very interesting if such correlations could be observed experimentally.  
 
 Finally, we note that the DPS HBT correlations are not limited to photon production. In particular these effects were not included in the CGC calculation of di-hadron production at forward rapidities \cite{dihadron}. Quite clearly a calculation similar to ours can be performed for double gluon 
 inclusive production and it should also lead to HBT correlations.  Indeed we expect that for gluons the suppression of the correlated part at high $k_T$ will be significantly smaller than for photons, since it involves production of only two high-$k_T$ particles in the final state, rather than four as in the present case.


\begin{acknowledgments}
 A.R. would like to thank the Particles, Astrophysics and Nuclear Physics Group in the University of Connecticut for the hospitality where part of this work were done.
 This research was supported in part by Conicyt (MEC) PAI 80160015.  The work of A. K. 
was supported in part by the NSF Nuclear Theory grant 1614640. The work of A. R. 
was supported in part by  Fondecyt grant 1150135, ECOS-Conicyt C14E01, Anillo ACT1406 and Conicyt PIA/Basal FB0821. 
\end{acknowledgments}


\end{document}